\documentclass[12pt,fleqn]{article}

\usepackage{amsmath,amssymb,graphicx}

\usepackage[a4paper,textwidth=15.9cm,textheight=23.2cm]{geometry}
\usepackage{latexsym}
\usepackage{cite}
\usepackage{bbm}
\usepackage{mathrsfs}

\numberwithin{equation}{section}

\def\AdSs5{$AdS_5$}
\def\AdSS5{$AdS_5$}
\def\AdS5s5{$AdS_5 \times S^5$}

\newcommand{\be}{\begin{equation}}
\newcommand{\ee}{\end{equation}}
\newcommand{\ba}{\begin{eqnarray}}
\newcommand{\ea}{\end{eqnarray}}
\newcommand{\bdm}{\begin{displaymath}}
\newcommand{\edm}{\end{displaymath}}

\newbox\SlashedBox
\def\fs#1{\setbox\SlashedBox=\hbox{#1}
\hbox to 0pt{\hbox to 1\wd\SlashedBox{\hfil/\hfil}\hss}{#1}}
\def\hboxtosizeof#1#2{\setbox\SlashedBox=\hbox{#1}
\hbox to 1\wd\SlashedBox{#2}}

\def\ms#1{\setbox\SlashedBox=\hbox{$#1$}
\hbox to 0pt{\hbox to 1\wd\SlashedBox{\hfil/\hfil}\hss}#1}

\def\t2{\tau_2}
\def\IZ{\relax\ifmmode\mathchoice {\hbox{\cmss Z\kern-.4em Z}}
{\hbox{\cmss Z\kern-.4em Z}} {\lower.9pt\hbox{\cmsss Z\kern-.4em
Z}} {\lower1.2pt\hbox{\cmsss Z\kern-.4em Z}} \else{\cmss
Z\kern-.4em Z}\fi}

\def\veps{\varepsilon}

\def\c1{{\chi^1}}

\def\N4{{\cal N}=4}
\def\half{\frac{1}{2}}

\DeclareMathAlphabet{\mathpzc}{OT1}{pzc}{m}{it}
\begin{document}


\thispagestyle{empty}
\renewcommand{\thefootnote}{\fnsymbol{footnote}}

{\hfill \parbox{4cm}{
        HU-EP-04/62}}

\bigskip\bigskip

\begin{center} \noindent \Large \bf
Strong Coupling Dynamics of the Higgs Branch:
\\ Rolling a Higgs by Collapsing an Instanton
\end{center}

\bigskip\bigskip\bigskip

\centerline{Zachary Guralnik\footnote[1]{
zack@physik.hu-berlin.de}}
\bigskip
\bigskip

\centerline{\it Institut f\"ur Physik} \centerline{\it
Humboldt-Universit\"at zu Berlin} \centerline{\it Newtonstra{\ss}e
15} \centerline{\it 12489 Berlin, Germany}
\bigskip

\bigskip\bigskip

\bigskip\bigskip

\renewcommand{\thefootnote}{\arabic{footnote}}

\centerline{\bf \small Abstract}
\medskip

{\small We construct the dual supergravity description of strongly
coupled, large $N$, eight-supercharge gauge theories with
fundamental hypermultiplets at points on the mixed Coulomb-Higgs
branch. With certain assumptions about unknown couplings of
D-branes to supergravity, this construction gives the correct
metric on the hypermultiplet (Higgs-branch) component the moduli
space, which decouples from the vector multiplet (Coulomb-branch)
moduli. Going beyond the geodesic approximation, we find that the
dynamics of a hypermultiplet VEV rolling towards a singularity on
the Higgs component of the moduli space is sensitive to the vector
multiplet moduli.  The dual description of the approach to the
singularity involves collapsing ``instantons'' of a non-Abelian
Dirac-Born-Infeld theory in a curved background. In general, we
find a decelerating approach to the singularity, although the
manner of deceleration depends on the vector multiplet moduli.
Upon introducing a potential on the Higgs branch of a four
dimensional ${\cal N}=2$ theory coupled to gravity, this
deceleration mechanism might lead to interesting inflating
cosmologies analogous those studied recently by Alishahiha,
Silverstein and Tong.}

\newpage

\section{Introduction}

The holographic relation between string theory and Yang-Mills
theories \cite{Maldacena:1997re,Gubser:1998bc,Witten:1998qj} known
as AdS/CFT duality has proven a powerful tool for studying
strongly coupled large $N$ Yang-Mills theories.  In its original
form, this duality was only applicable to a certain class of gauge
theories with matter in adjoint representations.  More recently,
there have been numerous studies of holographic dualities relating
string theory and gauge theories with fundamental representations.
This work has focused largely on electrically confining
``QCD-like'' theories. Instead of considering electrically
confining theories, we will construct backgrounds dual to points
on the mixed Coulomb-Higgs branch of eight-supercharge Yang-Mills
theories with fundamental representations, in various dimensions.
This work is a sequel to \cite{Guralnik:2004ve}, in which the AdS
dual of the Higgs branch of a four-dimensional ${\cal N}=2$ theory
was constructed.

The backgrounds we will describe can be used to study the strong
coupling, large $N$ dynamics on the Higgs component of the moduli
space. Some aspects of this dynamics are known exactly already,
since the two derivative effective action on the Higgs component
of the moduli space is not renormalized \cite{Argyres:1996eh}.
Realizing this non-renormalization theorem in the dual description
provides a test of AdS/CFT duality and constrains unknown terms in
the D-brane effective action, including non-minimal couplings to
bulk fields. In some situations however, the two derivative
effective action can not be expected to give a correct description
of the dynamics. One such situation is time dependent motion
towards a singularity of the moduli space.  We will study this
motion using the holographic dual description. Similar dynamics
has been studied in the context of the Coulomb branch of ${\cal
N}=4$ super Yang-Mills theory \cite{Kabat:1999yq}, and gives rise
to interesting inflating cosmologies upon introduction of a
potential on the Coulomb-branch and coupling to gravity
\cite{Silverstein:2003hf,Alishahiha:2004eh}(see also \cite{Chen}
for related work).

In \cite{Guralnik:2004ve}, the AdS description of the Higgs branch
of a four-dimensional ${\cal N}=2$ Sp(N) gauge theory with four
fundamental hypermultiplets and one anti-symmetric hypermultiplet
was constructed. This theory is conformal at the origin of moduli
space and is dual to string theory in the background $AdS^5 \times
S^5/Z^2$ with D7-branes wrapping the $Z^2$ fixed surface (an
O7-plane) \cite{Aharony:1998xz,Fayyazuddin:1998fb}, which has the
geometry $AdS^ 5 \times S^3$.  In \cite{Guralnik:2004ve}, it was
argued that the D7-brane equations of motion are solved by field
strengths which are self dual with respect to a flat four
dimensional metric,  despite the curved background and the
infinite number of higher dimension operators in the D7-brane
action, which includes non-minimal couplings to bulk fields. These
instanton solutions correspond to points on the Higgs branch, in
keeping with the known one-to-one map between the ADHM constraints
\cite{ADHM} used to construct instantons and the F and D-flatness
conditions which, modulo gauge transformations, determine the
Higgs branch \cite{Witten:1995gx,Douglas:1996uz} (see also
\cite{Dorey:2002ik} for a review). These instantons may be
interpreted as D3-branes dissolved within the D7-brane
\cite{Douglas:1995bn}.  Avoiding a back-reaction due to dissolved
instantons requires that the instanton number, or number of
dissolved D3-branes, is small compared to $N$.  For this reason,
the construction of \cite{Guralnik:2004ve} only applies to a
portion of the Higgs branch.

We will generalize the construction of \cite{Guralnik:2004ve}  to
include parts of the mixed Coulomb-Higgs branch in eight
supercharge theories in various dimensions. We will specifically
consider eight supercharge theories arising from systems of $N$ Dp
branes coincident with $N_f$ Dp+4 branes, with $p\le 3$. For
$N\rightarrow\infty$ with large 't Hooft coupling and $N_f$ fixed,
these theories are dual to string theory in the near horizon
geometry of the Dp-branes, with $N_f$ appropriately embedded probe
Dp+4 branes.

To two derivative order, the effective action on the Higgs branch
is given exactly by the tree level result \cite{Argyres:1996eh}.
Furthermore the corresponding metric on the moduli space is
equivalent to the metric which describes slowly varying
``instantons'' of super Yang-Mills theory in $p+5$ flat dimensions
\cite{Douglas:1996uz,Witten:1995gx,Dorey:2002ik}. At strong
coupling, AdS/CFT duality relates the effective action on the
Higgs branch to the dynamics of slowing varying instantons in the
Dp+4-brane theory, rather than flat space super Yang-Mills.
Nevertheless,  it was argued in \cite{Guralnik:2004ve} that (for
p=3) the result is the same at two derivative order, provided
certain constraints on unknown terms of the non-Abelian
Dirac-Born-Infeld action in a curved background are satisfied. In
fact, the non-renormalization of the metric on the Higgs branch
implies that these constraints must be satisfied.

The non-renormalization of the metric on the Higgs branch follows
from the local factorization the moduli space into a
hypermultiplet and a vector-multiplet component,  and the fact
that the gauge coupling can be viewed as a component of a
background vector multiplet \cite{Argyres:1996eh}.  We will see
that the factorization of the moduli space into Coulomb and Higgs
components is realized in the AdS description of the mixed
Coulomb-Higgs branch.

The low energy dynamics of the Higgs branch has properties which
may lead to interesting cosmologies upon introduction of a
potential and coupling to gravity.   These properties are similar
to those discussed in \cite{Silverstein:2003hf,Alishahiha:2004eh},
where a mechanism for slow roll inflation in the presence of a
steep potential was discovered in the context of a strongly
coupled gauge large N gauge theory, specifically a deformation of
${\cal N}=4$ super Yang-Mills coupled to gravity.  This mechanism
arises from the higher derivative structure of the effective
action on the Coulomb branch, which gives rise to an upper bound
on velocity in the moduli space.  The upper bound follows from
causality in the AdS description \cite{Kabat:1999yq}, and implies
that a certain time dependent Coulomb branch VEV rolling towards a
singularity in the moduli space will decelerate.  Upon suitably
deforming the theory and coupling to gravity, there are
cosmological solutions in which this time dependent modulus plays
the role of the inflaton.  A natural question is whether
deceleration is a generic feature of time dependent moduli
approaching a singularity in the moduli space.

Using the dual supergravity description,  we will study the
dynamics of time dependent Higgs branch moduli, keeping the
Coulomb branch moduli fixed. In the geodesic, or ``moduli space,''
approximation obtained by truncating to the two derivative
effective action, this dynamics is insensitive to the Coulomb
branch moduli and does not exhibit deceleration near
singularities. However, the geodesic approximation is not always
reliable, due to important corrections from integrating out states
which become light as one approaches a singularity. We will find a
deceleration mechanism for Higgs branch moduli approaching a
singularity.  The manner of deceleration depends on the
Coulomb-branch moduli, and does not always follow from causality
in the dual supergravity description.

We will focus on motion towards a singularity of the Higgs branch
which is dual to a shrinking instanton. No attempt will me made to
compute the full effective action including higher derivatives.
Instead we will consider time dependent solutions of the Dp+4
equations of motion which have instanton number one and possess
the usual symmetries of the static one-instanton solution. Since
the time dependent solutions are not self-dual,  it is not obvious
how to precisely identify the corresponding points on the Higgs
branch.  One way to do a more precise analysis would be to
integrate out fluctuations transverse to the instanton moduli
space (giving the effective action). Nevertheless, without exactly
identifying points in the moduli space, qualitative properities of
the evolution will be clear. The geodesic approximation, which is
given by the two derivative effective action, incorrectly predicts
that a shrinking instanton will collapse at a constant rate.
Although the Dp+4 equations of motion admit the usual static
instanton solutions, and the two derivative effective action on
the instanton moduli space is the same as that of flat space
Yang-Mills theory, the equations describing a collapsing instanton
differ from those of flat space Yang-Mills theory. Furthermore,
these equations depend on parameters dual to Coulomb-branch
moduli.

For vanishing Coulomb-branch moduli,  there are decelerating
solutions,  much like those discussed in
\cite{Kabat:1999yq,Silverstein:2003hf,Alishahiha:2004eh}, for
which deceleration follows from causality in the supergravity
description.   In this case it takes infinite time to reach the
singularity at the origin of the Higgs branch.  For certain
non-vanishing Coulomb moduli there is also a deceleration
mechanism.  However, once the Higgs VEV becomes small compared to
the Coulomb branch moduli,  deceleration takes a different form
which follows from particle production rather than a causal speed
limit in the dual description.  In this case deceleration is not
strong enough to prevent reaching the small instanton singularity
in a finite time. The rate of instanton collapse vanishes at the
time of collapse and the kinetic energy of the collapse gets
converted into the energy of modes transverse to the instanton
moduli space, which are dual to states which become light at the
origin of the Higgs branch. It is conceivable that, in a
cosmological setting, this could be interpreted as reheating.

The organization of this article is as follows.  In section 2, we
review basic properties of the Higgs branch of eight supercharge
theories arising from Dp-Dp+4 systems,  In section 3, we discuss
the supergravity description of a portion of the mixed
Coulomb-Higgs branch. In section 4 we show that this description
produces the correct metric on Higgs component of the moduli
space. In section 5, we use the supergravity description to study
time dependent solutions, going beyond the geodesic approximation.
In section 6, we conclude and discuss open problems.

\section{The Higgs branch of Dp -- Dp+4 systems}

We wish to study eight-supercharge Yang-Mills theories in p+1
dimensions corresponding to Dp--Dp+4 systems.  For $p<3$ we will
consider $N$ Dp-branes and $N_f$ Dp+4-branes,  such that the $p+1$
dimensional theory has $U(N)$ gauge symmetry and contains, besides
the vector-multiplet, one adjoint hypermultiplet and $N_f$
fundamental hypermultiplets. The fundamental hypermultiplets arise
from strings stretched between the Dp-branes and Dp+4-branes.
Specifically, we take the Dp+4-branes to extend in directions
$x^{0 \cdots p+4}$ and the Dp-branes to extend in directions
$x^{0\cdots p}$.  For $p=3$,  the gauge theory arising from the
D3-D7 configuration has a Landau pole,  reflecting an un-cancelled
tadpole in the string theory background.  This pathology seems to
avoided in the strict $N = \infty$ limit at finite $N_f$,  in
which the beta function for the 't Hooft coupling vanishes and the
back-reacted geometry involves a D7-brane wrapping a {\it
contractible} $S^3$ such that there is no net D7-charge
\cite{Karch:2002sh}. Alternatively one can consider the case in
which there are $4$ D7-branes coincident with an O7-plane, such
that one has a consistent tadpole free string background. This
yields a conformal gauge theory even at finite $N$ (see
\cite{Aharony:1998xz,Fayyazuddin:1998fb}) with Sp(N) gauge
symmetry and SO(8) flavour symmetry,  one hypermultiplet in the
anti-symmetric representation and four hypermultiplets in the
fundamental representation.

For definiteness, we will focus on the U(N) theories. The analysis
can be generalized to other cases. The superpotential is
\begin{align}
W= \tilde Q_i X Q_i + {\rm tr} X[Y,Z]\, , \end{align} where $Q_i$
and $ \tilde Q_i$ comprise fundamental hypermultiplets labeled by
the flavor index $i$, $Y$ and $Z$ belong to an adjoint
hypermultiplet, and $X$ belongs to the vector multiplet. There
must be at least two flavors for a Higgs branch to exist. We will
use lowercase letters to denote the bottom components of chiral
superfields.  On the Higgs branch, the vector multiplet moduli $x$
vanish while $q_i$ and $\tilde q_i$ have non-zero expectation
values. There are also mixed Coulomb-Higgs vacua, for which
$q_i,\tilde q_i$ and $x$ all have non-zero expectation values.

In the string theory realization of the theories which we
consider, the Dp+4-branes will always be taken to be coincident,
while the Dp-branes (before taking a near horizon limit) may be at
various locations in the transverse directions $x^{p+5} \cdots
x^9$, which correspond to various Coulomb-branch moduli. Mixed
Coulomb-Higgs vacua exist whenever some of the Dp-branes are
coincident with the Dp+4-branes.  We will specifically consider
the large $N$ limit in which in a fixed number $k$ of the
Dp-branes are coincident with the Dp+4-branes while the remaining
$N-k$ Dp-branes are a distance $M\alpha'$ from the Dp+4-branes.
This corresponds to vacua with
\begin{align}\label{Coul} x = \begin{pmatrix} M& & & & & \cr &\ddots & & & &  \cr
        & &M & & & \cr & & &0 & &  \cr & & & &\ddots & \cr
        & & & & & 0
\end{pmatrix}\, .
\end{align}
for which the F-flatness equations $\tilde q_i x = x q_i = 0$
permit fundamental hypermultiplet expectation values in which only
the last k entries of $q$ and $\tilde q$ are non-zero;
\begin{align}\label{cond}q_i = \begin{pmatrix} 0 \cr \vdots \cr 0 \cr \alpha_{i1} \cr
\vdots \cr \alpha_{ik} \end{pmatrix}\, \qquad  \tilde q_i =
\begin{pmatrix} 0 & \cdots & 0 & \beta_{i1} & \cdots
& \beta_{ik} \end{pmatrix}\, .\end{align} There are additional F
and D-flatness constraints which we have not explicitly written.

Nonzero entries in (\ref{cond}) physically correspond to Dp-branes
which are coincident with and dissolved within the Dp+4-branes.
Dissolved Dp-branes can be viewed as instantons in the $p+5$
dimensional world-volume theory on the Dp+4 branes
\cite{Douglas:1995bn}, due to the Wess-Zumino coupling
\begin{equation} S_{WZ} = T_{p+4}\, (2\pi\alpha')^2\, \int C^{(p+1)}  \wedge {\rm tr}
F\wedge F\, . \end{equation}  In fact, there is a well known exact
map between the moduli space of Yang-Mills instantons and the
Higgs branch of the $p+1$ dimensional theory arising on the Dp --
Dp+4 intersection.  The ADHM constraints from which instantons are
constructed \cite{ADHM} are equivalent to the F and D flatness
equations \cite{Witten:1995gx,Douglas:1996uz,Dorey:2002ik} of the
$p+1$ dimensional theory.

The metric on the Higgs branch is known to be tree level exact.
\cite{Argyres:1996eh}.  A brief summary of the argument behind
this non-renormalization is as follows.  The Kahler potential is a
function $K(h_i,\tilde h_i, \phi, h_i^{\dagger}, {\tilde
h}_i^\dagger, \phi^\dagger)$ of the hypermultiplet moduli,
collectively denoted by $h_i, \tilde h_i$, and the vector
multiplet moduli $\phi$. Lorentz invariance together with ${\cal
N}=2$ supersymmetry requires that $K= K_H(h_i,h_i^\dagger) +
K_V(\phi,\phi^\dagger)$, such that the moduli space is locally a
product of Higgs-branch and Coulomb-branch components.  Finally,
the gauge coupling can be viewed as a component of a background
vector multiplet, on which $K_H$ does not depend.

For the theories associated with the Dp-Dp+4 system, the
Hyperk\"ahler metric on the Higgs branch  is equivalent to that
which descibes slowly varying ``instantons'' in eight dimensional
super Yang-Mills theory
\cite{Douglas:1996uz,Witten:1995gx,Dorey:2002ik}. These instantons
correspond to field strengths which are self-dual with respect to
the flat metric in the directions $x^{4,5,6,7}$. The effective
action for slowly varying instantons takes the form,
\begin{align} \int dx^0 \cdots dx^3 G_{ij}({\cal
M})\partial_{\mu}{\cal M}^i \partial^\mu {\cal M}^j \end{align}
where ${\cal M}^i$ are the instanton moduli, which are allowed to
depend on the coordinates $x^{0,1,2,3}$, and $G_{ij}$ is the
metric on the moduli space.

\section{Supergravity description of the Higgs branch}

For $N \rightarrow \infty$ with  $N_f$ fixed and 't Hooft coupling
$\lambda =g^2N
>>1$, the super Yang-Mills theory associated with the Dp-Dp+4
system is dual to supergravity in the near horizon geometry of the
N Dp-branes, with the Dp+4 branes  treated as probes. The near
horizon geometry of the Dp-branes is
\begin{align}\label{geom}
&ds^2 =   \alpha'\left[H(r)^{-1/2} \eta_{\mu\nu}dx^\mu dx^\nu +
H(r)^{1/2}
\sum_{a=1}^{9-p}(dy^a)^2 \right] \nonumber \\
&C_{01\cdots p}  =  {\alpha'}^2H(r)^{-1} \\
&e^{-2(\phi-\phi_\infty)} = {\alpha'}^{3-p} H(r)^{(p-3)/2}\,
,\qquad e^{\phi_{\infty}}= g_s\, ,\nonumber
\end{align}
where $\mu = 0\cdots p$, $\eta_{\mu\nu}$ is the $p+1$ dimensional
minkowski metric, and
\begin{align} r^2 &=
\sum_{a=1}^{9-p}(y^a)^2\, , \qquad H(r) =
\frac{d_p\lambda}{r^{7-p}},\qquad d_p =
2^{7-2p}\pi^{\frac{9-3p}{2}}\Gamma(\frac{7-p}{2}), \qquad  \nonumber \\
\lambda &= (2\pi)^{p-2}{\alpha'}^{(p-3)/2}g_sN\, .\end{align} This
supergravity background is reliable (see \cite{Itzhaki:1998dd}) in
the region
\begin{align} \left(\lambda N^{\frac{4}{p-7}}
\right)^\frac{1}{3-p}<< r << \lambda^{\frac{1}{3-p}}\,
.\end{align}

The Dp+4 probes which gives rise to hypermultiplets with mass $M$
are localized in the geometry (\ref{geom}) at
\begin{align}\label{embed} &y^{a} = 0\,\,\,\, {\rm for}\,\,\,\, 5\le a \le
8-p, \nonumber\\ &y^{9-p} = M \, .
\end{align}
There must be at least two Dp+4 branes, corresponding to a minimum
of two flavors, for a Higgs branch to exist.  The number of
Dp+4-branes must be held fixed in the large $N$ limit to avoid
consideration of the back reaction.  The induced metric on the
Dp+4-branes is
\begin{align}\label{indmet}
ds^2 = \alpha'\left[H^{-1/2}(r) \eta_{\mu\nu}dx^\mu dx^\nu +
H^{1/2}(r) dy^m dy^m\right]\, ,
\end{align} with $m=1\cdots 4$ and $r^2 = y^m y^m + M^2$.
For $p=3$ this geometry  is $AdS^5 \times S^3$ if $M=0$, or
asymptotically $AdS^5 \times S^3$ when $M\ne 0$. The stability of
such embeddings was discussed in \cite{Karch:2002sh}, where it was
shown that the Breitenlohner-Freedman bound is satisfied
(saturated in fact), while the supersymmetry of the embedding was
explicitly shown in \cite{Skenderis:2002vf}.  For $p<3$, we expect
that the embedding (\ref{embed}) remains supersymmetric although,
to our knowledge, this has not been explicitly demonstrated.

The geometry associated with the mixed Coulomb-Higgs branch vacua
(\ref{Coul}),(\ref{cond}) is also (\ref{geom}) with the embedding
(\ref{embed}), provided that the number, $k$, of vanishing
diagonal entries in (\ref{Coul}) is fixed in the large $N$ limit.
This corresponds to a configuration of $k$ Dp-branes coincident
with Dp+4-branes, in the near horizon geometry (\ref{geom})
arising from $N-k$ Dp-branes at a non-zero distance from the
Dp+4-branes. (\ref{embed}). At generic points in the Higgs-branch
component of the moduli space, the k Dp-branes are dissolved in
the Dp+4-branes. The effective action of the $Dp+4$ branes is a
$p+5$-dimensional non-Abelian Dirac-Born-Infeld theory in a curved
space. Nevertheless, the correspondence between instantons and the
Higgs branch leads one to expect that conventional Yang-Mills
instantons should also solve the Dp+4 equations of motion.

The Dp+4 brane action takes the form
\begin{align}\label{ac}
S &= T_{p+4} \int  \sum_r C^{(r)} \wedge{\rm tr} e^{2\pi  \alpha'
F}\wedge\left(\frac{\hat{\cal A}(4\pi^2  \alpha' R_T)}{{\hat{\cal
A}(4\pi^2  \alpha' R_N)}}\right)^{1/2}\nonumber\\
&- T_{p+4}\int\,d^{p+5}\xi\, \sqrt{g} e^{-(\phi-\phi_\infty)}(2\pi
\alpha')^2\frac{1}{2}{\rm tr}\left( F_{AB}F^{AB} \right) + \cdots
\, ,
\end{align} where we have not written terms involving fermions and scalars.
The action (\ref{ac}) is the sum of a Wess-Zumino
term\footnote{The ``A-roof genus'' ${\hat {\cal A}}$ appearing in
the Wess-Zumino term (see \cite{Green:1996dd}) has an expansion in
terms of even powers of $R_T$ and $R_N$, which are pull-backs of
the Reimann curvature two-form to the tangent and normal bundle
respectively.}, a Yang-Mills term, and an infinite number of
corrections at higher orders in $\alpha'$, indicated by $\cdots$
in (\ref{ac}). Even in flat space, little is known about these
corrections in the non-Abelian case, with the exception of a few
of the leading order terms. Even less is known about higher
dimension non-minimal couplings of D-brane gauge fields to bulk
fields such as the curvature, Ramond-Ramond field strengths, and
derivatives of the dilaton.

Let us now consider turning on gauge fields $A_m(y^n)$, where the
indices $m,n=1\cdots 4$ are associated with the coordinates $y^m$
appearing in the induced metric (\ref{indmet}) on the Dp+4-brane.
At leading order in the $\alpha'$ expansion,  field strengths
$F_{mn}$ which are self-dual with respect to the {\it flat} four
dimensional metric $ds^2 = dy^mdy^m$ solve the equations of motion
due to a conspiracy between the Wess-Zumino and Yang-Mills term.
Inserting the explicit form of the induced metric, Ramond-Ramond
potential, and dilaton into the leading terms of the Dp+4 action
(\ref{ac}) gives,
\begin{align}\label{sqr}
S &= \frac{N}{(2\pi)^6 \lambda} \int\, d^{p+1}x\,d^4y\, H(r)^{-1}
\, {\rm
tr}\left(F_{mn}F_{mn}-\frac{1}{2}\epsilon_{mnrs}F_{mn}F_{rs}
\right)
 \nonumber \\
&= \frac{N}{2(2\pi)^6\lambda} \int \, d^{p+1}x\,d^4y\,  H(r)^{-1}
{\rm tr}{F_-}^2 \, ,
\end{align}
where $r^2 = \sum_{m=1}^4 {(y^m)}^2 + M^2,\,\, F_- =
\frac{1}{2}(F_{mn} - \frac{1}{2}\epsilon_{mnrs}F_{rs})$. We have
written the Dp+4-brane tension explicitly in terms of $\alpha'$
and parameters of the dual $p+1$ dimensional gauge theory:
\begin{equation}T_{p+4} = \frac{1}{g_s (2\pi)^{p+4}
{\alpha'}^{(p+5)/2}} = \frac{N}{(2\pi)^6 {\alpha'}^4 \lambda}\,
,\end{equation} where $\lambda = g^2_{YM} N$ is the 't Hooft
coupling.  Field strengths for which $F_- = 0$ are self-dual with
respect to the flat metric $dy^mdy^m$ and manifestly solve the
equations of motion.

The self dual field strengths correspond to points on the Higgs
component of the moduli pace, while the parameter $M$ is a
Coulomb-branch modulus.  Note that the same self-dual solutions
exist even if one does not take the near horizon limit, such that
$H(r) = {\alpha'}^2 + d_p\lambda/r^{7-p}$. The usual flat space
instantons continue smoothly into the near horizon region. Thus we
expect that the the map between instanton moduli in the
supergravity description and points on the Higgs component of the
moduli space is the usual one. Strictly speaking, we can only
describe a portion of the Higgs branch in this way, since the
number of dissolved D3-branes is held fixed in the large $N$ limit
to avoid having to consider back reaction.

The exact correspondence between the Higgs branch and Yang-Mills
instantons leads one to expect that self-dual field strengths
should remain solutions of the Dp+4-brane action to all order in
the $\alpha'$ expansion. This implies constraints on various
unknown terms of the non-abelian Dirac Born Infeld action in a
curved background (see \cite{Guralnik:2004ve}).  These constraints
are somewhat beside the point of this paper, but are worth
mentioning briefly. Among the constraints is one which applies to
terms quadratic in the world-volume field strengths, of the
general form $f(R, {\cal F}^{(p+2)}, \phi) F^2$. Here $f(R, {\cal
F}^{(p+2)}, \phi)$ depends on the pull-backs of the bulk fields
which are nonvanishing in the background (\ref{geom}), as well as
their derivatives. We have not written any explicit contractions
of Lorentz indices, of which there are numerous possibilities.  At
order $\alpha'$, terms of the form $R^2F^2$ have been discussed in
\cite{Frey:2003jq,Wijnholt:2003pw}. Assuming $F_- =0$ connections
solve the equations of motion, there are instanton solutions for
which the field strength is locally arbitrarily small compared to
the bulk fields, and one can expand in the field strength.  At
quadratic order in field strengths the CP odd Wess-Zumino term
proportional to $\int H(r)^{-1} F^*F$ is exact and receives no
corrections at any order in $ \alpha'$. In order to preserve the
$F^- = 0$ solutions, the quadratic CP even term must be of the
form $\int H(r)^{-1} F^2$ with exactly the same coefficient. As
discussed above, this is already the case at leading order in
$\alpha'$. Thus at every higher order in $ \alpha'$, the terms of
the form ${\alpha'}^n f_n(R, {\cal F}^{(p+2)}, \phi) F^2$ must sum
to zero when the bulk fields are set to the background values
(\ref{geom}). These constraints are very similar to constraints on
the non-Abelian Dirac-Born-Infeld action in flat space which have
been found by requiring that stable-holomorphic bundles solve the
equations of motion \cite{Koerber:2002zb}.

Assuming that terms of the form $f(R, {\cal F}^{(p+2)}, \phi) F^2$
collectively sum to zero, the action to order ${\alpha'}^2$ is
given by
\begin{align}\label{ffour} S = &T_{p+4} \int C^{(p+1)}\wedge F\wedge F \nonumber \\
& + T_{p+4} \int \, \sqrt{-g}e^{-(\phi-\phi_\infty)}
{\rm tr}\left[F^{AB}F_{AB} + \alpha'^2\left(\frac{1}{24}
F_{AB}F^{BC}F_{CD}F^{DA}
\right.\right. \\
&\left.\left. +
\frac{1}{12}F_{AB}F^{BC}F^{DA}F_{CD}-\frac{1}{48}F_{AB}F^{BA}F_{CD}F^{DC}
-\frac{1}{96}F_{AB}F_{CD}F^{BA}F^{DC}\right)\right]\, \nonumber ,
\end{align} where the $F^4$ terms were computed in
\cite{Gross:1986iv,Tseytlin:1986ti,Tseytlin:1997cs,Bergshoeff:2001dc}.
For the embedding (\ref{embed}) in the background (\ref{geom}),
with non-zero field strengths in the $y^m$ directions only, this
action becomes
\begin{align}\label{fsq}
S = \frac{N}{2(2\pi)^6 \lambda}\int\, d^px\,d^4y\,\, {\rm tr}\,
&\left[ H^{-1}(r) F^-_{mn}F^-_{mn}\right. \nonumber
\\
 -&\frac{1}{12}H^{-2}(r)\left.
\left(2 F^+_{mn}F^+_{mn}F^-_{rs}F^-_{rs} +
F^+_{mn}F^-_{rs}F^+_{mn}F^-_{rs} \right)\right]\, ,
\end{align}
where $r^2 = y^my^m + M^2$, and we have written $F_{mn}$ in terms
of self-dual and anti-self-dual components, $F^\pm_{mn}\equiv
\half\left(F_{mn}\pm \half\veps_{mnrs}F_{rs}\right)$. Since all
terms in (\ref{fsq}) involve two factors of $F^-$, field strengths
for which $F^- =0$ are manifestly solutions. In light of the
correspondence between instantons and the Higgs branch, this
should hold for the full Dirac-Born-Infeld action.

\section{The metric on the Higgs Branch}

To two derivative order, the effective action on the Higgs branch
of the eight supercharge $p+1$ dimensional theory is tree level
exact, and is known to be equivalent to the action describing the
geodesic approximation \cite{Manton:1981mp} for slowly varying
``instantons'' in a $p+5$ dimensional Yang Mills theory.  These
instantons are localized in the directions $y^m \sim x^{p+1}
\cdots x^{p+4}$ and, in the static case, have no dependence on the
coordinates $x^\mu \sim x^0 \cdots x^p$. The associated gauge
connections have the form $A_m = A_m^{\rm inst}(y^n, {\cal M}_i)$,
where ${\cal M}_i$ are the instanton moduli.  There are slowly
varying approximate solutions in which the moduli ${\cal M}_i$
depend on the coordinates $x^\mu$.  The effective action governing
these moduli has the form
\begin{equation}
S= \int d^{p+1}x G_{ij}({\cal M})\partial_\mu {\cal M}^i
\partial^\mu {\cal M}^j
\end{equation}
and is equivalent to the two-derivative effective action on Higgs
the branch,  where the ${\cal M}_i$ are interpreted as scalars
parameterizing the Higgs-branch.

In the dual description of the $p+1$ dimensional theory given by
the geometry (\ref{geom}) with Dp+4-branes embedded according to
(\ref{embed}), one can compute the metric on the Higgs branch by
finding the action for slowly varying instantons on the
Dp+4-branes.  In spite of the curved background and infinite
number of higher dimension operators in the Dp+4-brane action, the
result must be identical to that obtained from conventional
Yang-Mills theory in $p+5$ flat dimensions.

The metric can be extracted from the equations which
configurations of the form
\begin{align}\label{config} A_m = A^{inst}_m(y^n, {\cal M}_i(x^\alpha))\, ,
\qquad A_\mu = \Omega_i \partial_\mu  {\cal M}_i(x^\alpha)
\end{align} must satisfy in order to solve the equations of motion
at leading order in a $\partial_\mu$ expansion (see
\cite{Dorey:2002ik} for a review).  For this purpose, the relevant
terms in the Dirac Born Infeld action are those involving two
Greek indices. To order ${\alpha'}^2$, which becomes ${\cal
O}(1/\lambda)$ in the background we consider,  the terms of this
type arising from the action (\ref{ffour}) are,
\begin{align}\label{met}
S = &\frac{N}{(2\pi)^6\lambda}  \int\, d^{p+5} \xi\,\frac{1}{4}
{\rm tr}(F_{\mu m} F_{\mu m})  \nonumber\\
\,\,\, &+ \frac{N}{(2\pi)^6 \lambda}\int\, d^{p+5} \xi\, H(r)^{-1}
\frac{1}{12}{\rm tr} \left[F_{s\mu}F_{\mu
n}\left(\{F_{nr},F_{rs}\}
-\frac{1}{2}\delta_{sn}F_{tu}F_{ut}\right)\right. + \nonumber\\
&\qquad \qquad\frac{1}{2}\left. \left(F_{\mu
n}F_{nr}F_{s\mu}F_{rs} + F_{\mu n}F_{rs}F_{s\mu}F_{nr} -
\frac{1}{2}F_{\mu n}F_{rs}F_{n\mu}F_{sr}\right)\right] + \cdots
\,.
\end{align}
Note that the term quadratic in $F$ is just that of Yang-Mills
theory in $p+5$ flat dimensions; factors of the harmonic function
$H(r)$ appearing in the metric (\ref{geom}) cancel in this term.
This is the leading term in a large $\lambda$ expansion.  Since
$H^{-1} \sim r^{7-p}/\lambda$, the $F^4$ terms in (\ref{met}) are
subleading. However, the leading term should be the only term
contributing to the metric on the Higgs branch, since flat space
Yang-Mills gives the exact metric.

To see that the subleading term does not contribute, it is useful
to rewrite it in terms of the self-dual and anti-self-dual parts
of $F_{mn}$, giving
\begin{align}\label{metact} S&= \frac{N}{(2\pi)^6 \lambda} \int d^{p+5}\xi\,
\frac{1}{4}
{\rm tr}(F_{\mu m} F_{\mu m})  \nonumber\\
&+ \frac{N}{(2\pi)^6 \lambda}\int\, d^{p+5}\xi\,
H(r)^{-1}\,\frac{1}{48}{\rm tr} \left[F_{s\mu}F_{\mu
n}\left(\{F^+_{nr},F^-_{rs}\}+\{F^-_{nr},F^+_{rs}\}
\right)\right. + \nonumber\\
&\qquad \frac{1}{2}\left. \left(F_{\mu n}F^+_{nr}F_{s\mu}F^-_{rs}+
F_{\mu n}F^-_{nr}F_{s\mu}F^+_{rs} + F_{\mu
n}F^+_{rs}F_{s\mu}F^-_{nr}+ F_{\mu n}F^-_{rs}F_{s\mu}F^+_{nr}
\right)\right]
\end{align}
Hence, the subleading terms vanish when $F_- =0$.  The calculation
of the metric goes through as it does in flat space Yang-Mills.
For the configuration (\ref{config}), the equations of motion
$\frac{\delta}{\delta A_\mu} S =0$ become
\begin{align}\partial_\mu {\cal M}_i\left(D_m \frac{\delta A_m}{\delta {\cal M}_i} - D_m D_m
\Omega_i\right) = 0\, ,\end{align} which has a solution for
$\Omega_i$ as a function of $y^m$ and ${\cal M}_i$. Taking this
$\Omega_i$ and inserting (\ref{config}) into the action
(\ref{metact}) gives the metric on the Higgs branch;
\begin{align}
\partial_\mu{\cal M}^i \partial_\mu {\cal M}^j G_{ij}({\cal M}) =
\frac{N}{(2\pi)^6 \lambda}  \int d^4 y\,\frac{1}{4} {\rm
tr}(F_{\mu m} F_{\mu m})\, .
\end{align}
Note that the metric is insensitive to the embedding parameter $M$
which enters the action through the harmonic function $H(r)$ with
$r^2 = y^m y^m + M^2$.  This parameter corresponds to an
expectation value for a vector multiplet scalar. Insensitivity to
$M$ reflects the fact that the moduli space factorizes locally
into Higgs and Coulomb branch components \cite{Argyres:1996eh}.

In fact, all higher order contributions in the strong coupling
expansion of the metric must vanish, since the flat space
Yang-Mills result for the metric is exact.   Note that we have
obtained the correct metric at next to leading order without
considering possible non-minimal couplings of D-brane fields to
bulk fields.  The absence of a contribution to the metric from
these couplings yields another constraint on unknown bulk to brane
couplings.

\section{Time dependent solutions: beyond the geodesic
approximation}

We have argued that the AdS description of mixed Coulomb-Higgs
vacua gives the correct two-derivative effective action on the
Higgs component, corresponding to the geodesic motion of
instantons in a flat space Yang-Mills theory.  This result is
insensitive to the expectation values of vector multiplet scalars,
reflecting the local factorization of moduli space into Higgs and
Coulomb parts.  Nevertheless, we will see that the dynamics of a
time dependent Higgs field rolling towards a singular point the
moduli space is not necessarily described by the geodesic
approximation, nor is it independent of the Coulomb branch
expectation values.

At strong coupling, one particular approach to a singularity of
the Higgs branch can be studied by considering the equations of
motion of a collapsing instanton on the Dp+4 branes embedded in
the geometry (\ref{geom}) according to (\ref{embed}).    In
principle, the AdS description allows one to compute the low
energy effective action on the Higgs branch at strong 't Hooft
coupling, including higher derivative terms,  by integrating out
modes transverse to the instanton moduli space.  We leave this
computation for the future.  Instead,  we will look for
qualitative properties of approach to the singularity by examining
solutions of the Dp+4-brane equations of motion, without
integrating out modes transverse to the instanton moduli space.
Note that it does not make sense to integrate out these modes if
there is production of particles (W-bosons and their
superpartners) which become light as a Higgs scalar approach the
singularity, in which case the effective action on the Higgs
branch is not a good description. We will see that this
circumstance arises when the Coulomb branch modulus $M$ is
non-zero.

Exact collapsing instanton solutions are not self-dual and are
therefore not described by the ADHM construction. Without
integrating out modes transverse to the instanton moduli space, it
is not obvious how to map points on the Higgs branch, unless one
works in an approximation in which the deviation from self-duality
is small. For non-zero values of the Coulomb branch modulus $M$
and sufficiently small instanton size, we will be able to borrow
results obtained in \cite{Bizon:2003kz}, in which such an
approximation was used to describe collapsing instantons of
Yang-Mills theory in five-dimensional Minkowski space-time. For
vanishing Coulomb-branch moduli, we will find approximate
collapsing instanton solutions which are rather far from
self-duality. In this case we will only be able to make
qualitative statements about the behavior of the Higgs expectation
value.

\subsection{Causality in AdS}

As discussed in
\cite{Kabat:1999yq,Silverstein:2003hf,Alishahiha:2004eh}, the
dynamics of Coulomb-branch moduli in large N, strongly coupled,
${\cal N}=4$ super Yang-Mills theory is constrained by causality
in the dual AdS description.  In particular there is an upper
bound on the velocity with which moduli can roll towards a
singularity, which decreases with proximity to the singularity.
One might expect similar constraints to hold for the Higgs branch
(or mixed Coulomb-Higgs branch) of the eight supercharge theories
we have been considering. In this setting, the relevant part of
the induced metric on the Dp+4-branes is
\begin{align}\label{Dp4metric}
ds^2 =   \alpha'\left[- H(r)^{-1/2} dt^2 + H(r)^{1/2}
\sum_{m=1}^{4}dy^mdy^m \right]\, ,
\end{align}
with $r^2 = y^my^m+M^2$.  For $M=0$,  the point $y^m=0$
corresponds to the horizon $r=0$. Causality implies that it takes
a pointlike object infinite coordinate time $t$ (corresponding to
time in the dual $p+1$ dimensional gauge theory) to reach $y^m
=0$.  For $M\ne 0$, the point $y^m=0$ is not at the horizon and
causality does not prevent pointlike objects from reaching this
point in finite time. Although an instanton is not a pointlike
object, it seems a natural supposition that the size of an
instanton centered at the origin is also subject to causality
constraints. For $M=0$, one would expect that it takes infinite
time $t$ for an instanton to collapse, since the zero size
instanton is localized at the horizon. We will see this
expectation born out by some approximate collapsing instanton
solutions.  This behavior is very different from what is predicted
by the geodesic approximation (two-derivative effective action)
which gives collapse at a constant rate: $\ddot\rho =0$ where
$\rho$ is the instanton size.

For $M\ne 0$, causality can not rule out collapse in finite time.
Nevertheless, we will still find a deceleration mechanism which
violates the geodesic approximation.  This deceleration is
unrelated to causality in the dual supergravity description, and
is not strong enough to prevent collapse in a finite time. However
the rate of collapse does vanish in the zero size limit.

\subsection{Collapsing instantons}

The action for the Dp+4 brane with the embedding (\ref{embed}) is
\begin{align}\label{strx} S = &\frac{N}{(2\pi)^6\lambda}
\int d^{p+5} \xi\, {\rm tr}\left(\frac{1}{2} H(r)^{-1}
F^-_{mn}F^-_{mn}  + 2F_{m \mu}F_{m \mu} +
H(r)F_{\mu\nu}F_{\alpha\beta}\eta^{\mu\alpha}\eta^{\nu\beta}\right)\nonumber\\
&+ \cdots\, ,
\end{align}
where Greek indices indicate the coordinates $x^{0\cdots p}$ and
Roman indices indicate the coordinates $y^{1\cdots 4}$.  The
higher dimension terms indicated by $\cdots$ involve arbitrarily
high powers of $H(r)$ or $H(r)^{-1}$, depending on the numbers of
Greek or Roman indices.  Since $H(r) = \lambda/r^{7-p}$, a strong
coupling expansion becomes manifest only after the rescaling
$x^\mu \rightarrow \sqrt{\lambda} x^\mu, A_{\mu}\rightarrow
A_{\mu}/\lambda$. After this rescaling, the supergravity metric
has an overall $\sqrt{\lambda}$ in front, and (\ref{strx}) becomes
\begin{align}\label{ovrl}
S &= \frac{N}{(2\pi)^6 \lambda^2} \int d^{p+5} \xi\, {\rm
tr}\left(\frac{r^{7-p}}{2d_p} F^-_{mn}F^-_{mn}+2F_{m \mu}F_{m
\nu}\eta^{\mu\nu}+\frac{d_p}{r^{7-p}}
F_{\mu\nu}F_{\alpha\beta}\eta^{\mu\alpha}\eta^{\nu\beta}\right)
\nonumber\\ & + {\cal O}(\frac{1}{\lambda^3}),
\end{align}
where $r^2 = y^my^m +M^2$.  We will only consider the leading term
in the strong coupling expansion.

Since we focus on the quadratic part of the action, the reader
might incorrectly suspect that what we are doing is very different
from that of
\cite{Kabat:1999yq,Silverstein:2003hf,Alishahiha:2004eh}, where
behavior outside the geodesic approximation followed from
considering the full Dirac-Born-Infeld action.  In strongly
coupled large $N$ ${\cal N}=4$ super Yang-Mills theory,  the
dynamics of a particular Coulomb-branch modulus rolling towards a
singularity of the moduli space is described by the effective
action of a probe D3-brane in AdS,
\begin{align} \label{Coulomb} S= \int d^4 \xi X^4\left(\sqrt{1 -
\frac{\lambda{\dot X}^2}{X^4}} -1\right)\, ,
\end{align} with the singularity lying at $X=0$. The geodesic
approximation is $S \sim \int {\dot X}^2$,  which is blind to the
causal constraint $\sqrt{\lambda}\dot X \le X^2$ arising from an
infinite number of higher derivative corrections. It must be
remembered that in the case of the Higgs branch, we have not yet
integrated out modes which are transverse to the moduli space,
after which one would also obtain an action with an infinite
number of higher derivatives. Furthermore the action
(\ref{Coulomb}) also gives a leading contribution in a strong
coupling expansion (seen more clearly after rescaling $x^\mu
\rightarrow \sqrt{\lambda}x^\mu$).

To focus on time dependent solutions of (\ref{ovrl}) with no
dependence on $x^{1\cdots 3}$, it is sufficient to consider the
five dimensional action
\begin{equation}\label{unusualaction} S = \int dt\,d^4y\, {\rm tr}\left[F_{0m}F_{0m} -
\frac{1}{2d_p}{(y^my^m+M^2)}^{(7-p)/2}F^-_{mn}F^-_{mn}\right]
\end{equation}
This action has the usual static instanton solutions.  The SU(2)
single instanton solutions are
\begin{align}
A^a_{m}(y) = \frac{2\eta^a_{mn}(y^n - Y^n)}{(y- Y)^2 + \rho^2}
\end{align}
where  $m=1,2,3,4$, $a$ is a Lie algebra index, and $\eta^a_{nm}$
is the 't Hooft tensor;
\begin{align}
\eta^a_{mn} = -\eta^a_{nm},\, \eta^a_{ij} = \epsilon_{aij},\,
\eta^a_{i4} = \delta_{ia}\,\, {\rm with}\, i=1,2,3
\end{align}
We will study time-dependant solutions with which preserve the
symmetries of the single $SU(2)$ instanton centered at the origin
and have instanton number one. Unlike the flat space Yang-Mills
theory, the origin is a special point due to the curved geometry
of the Dp+4-brane embedding.  Note that this fact is not visible
in the metric on the instanton moduli space. We make the time
dependant ansatz
\begin{align}\label{ansatz} A^a_m &= \eta^a_{nl}x^l
\frac{{f}(\upsilon,t)}{\upsilon^2} \nonumber \\ A^a_0 &=
0\, ,
\end{align}
where $\upsilon^2 \equiv y^my^m$.  The Gauss law constraint,
$\frac{\delta}{\delta A_0} S =0$ is satisfied by this ansatz.  The
remaining equations of motion give\footnote{Some useful identities
involving the 't Hooft tensor are written in the appendix.};
\begin{align}\label{EOM}
\left(-\partial_t^2  + r^{7-p}
\partial_{\upsilon}^2 + \left(r^{7-p} + (7-p) \upsilon^2
r^{5-p}\right)\frac{1}{\upsilon}\partial_\upsilon \right)f
\nonumber
\\+ \frac{r^{7-p}}{\upsilon^2}(-4f + 6 f^2 -2f^3) + r^{5-p}(f^2
-2f) =0\, ,
\end{align}
with $r^2 = \upsilon^2 + M^2$.  We have rescaled coordinates to
remove numerical factors of $d_p$.

\subsection{Rolling Higgs at the origin of the Coulomb branch}

Let us now examine solutions of (\ref{EOM}) in the special case
$M^2 =0$ corresponding to the origin of the Coulomb branch.  For
$M^2 = 0$, (\ref{EOM}) becomes
\begin{align}\label{EOM2}
\left(-\partial_t^2  + \upsilon^{7-p}
\partial_{\upsilon}^2 +
(8-p)\upsilon^{7-p}\frac{1}{\upsilon}\partial_{\upsilon} \right)f
+ \upsilon^{5-p}Q(f) = 0\, ,\nonumber \\{\rm with}\,\,\,\, Q(f)
\equiv -18f +  13f^2 - 2f^3 + p(2f-f^2)
\end{align}
The usual static instanton solution is,
\begin{equation} f = \frac{2r^2}{r^2+\rho^2}\, ,
\end{equation}
which interpolates between the zeros $f=0$ and $f=2$ of $Q(f)$. As
long as a configuration $f(r,t)$ interpolates between these points
at any fixed value of $t$, the instanton number
$\frac{1}{16\pi^2}\int F \wedge F$ is equal to one. It is not
difficult to see that exact time dependent solutions can not be
self-dual. Thus there is no exact solution of the form  $f=
2r^2/(r^2+ \rho(t)^2)$,   although this may be taken as the
leading term in  an expansion valid for sufficiently slowly
varying instanton size. While this form is suitable for
determining the two derivative effective action, studying the
corrections to the geodesic approximation requires considering
configurations which are not self-dual.

The map between the time dependent instanton solution and points
on the Higgs branch is less clear than in the static case because
the solutions are not described by the ADHM construction. One way
to define the instanton size $\rho(t)$ is by the
moment\footnote{This is equal to the hyperk\"ahler potential
which, for a single static instanton localized at the origin,  is
proportional to the square of the instanton size} $\rho^2 \sim
\frac{1}{16\pi^2}\int r^2 F\wedge F$. Alternatively one can define
$\rho(t)$ as the value of $r$ for which $f(r,t)=1$. We do not
expect either of these definitions to map to points on the Higgs
branch in the same way as the instanton size in the static case.
However we will only be interested in qualitative behavior for
which either definition can be taken as a crude measure of the
dual Higgs VEV. Yet another possibility, which is more natural
from the point of view of the usual AdS/CFT dictionary, is to
define $\rho$ by the large $r$ asymptotics.  In the static case,
the leading large $r$ behavior of the field strength is $F^a_{mn}
\sim 4\rho^2 \eta^a_{mn}/r^4$. However this definition will not
prove useful for the approximate solution below,  which is only
valid at small $r$.

It is convenient to define $z\equiv \frac{2}{p-5} r^{(p-5)/2} $
such that (\ref{EOM2}) becomes
\begin{align}\label{EOM3}
\left(-\partial_t^2 + \partial_z^2 +
\frac{9-p}{p-5}\frac{1}{z}\partial_z \right)f +
\frac{1}{z^2}\frac{4}{(p-5)^2} Q(f) = 0
\end{align}
Equation (\ref{EOM3}) has approximate collapsing instanton
solutions;
\begin{equation}
f(z,t) \approx \tilde f(z-t)\, ,\end{equation} where $\tilde f(z)$
interpolates between $f=0$ and $f=2$ in a region of sufficiently
large $z$ (small r) such that $Q(\tilde f)/z^2 <<
\partial_z^2 \tilde f$ and $\partial_z \tilde f/z << \partial_z^2
\tilde f$.  This approximation is a good description of the
instanton collapse as $t$ increases,  but is not valid for
sufficiently early times (negative $t$), since the conditions $Q(
f)/z^2 << \partial_z^2 f$ and $\partial_z  f/z << \partial_z^2 f$
are violated. In terms of the coordinate $z$, the point at which
$f=1$ propagates with velocity $\dot z=1$.  Thus it takes infinite
time $t$ for the instanton to collapse,  which occurs when
$z|_{f=1} =\infty$.  This solution corresponds to a spherical
shell of topological charge moving at the speed of light.
Restricting to fixed points $y^m/|y|$ on the sphere, the metric
(\ref{Dp4metric}) is (for $M=0$) conformally equivalent to $dt^2 -
dz^2$.

The motion $\dot z =1$ of the collapsing instanton suggests a
rolling Higgs
\begin{equation}\label{decel} \langle \tilde q q \rangle \approx r^2\,\,\,
{\rm with} \,\,\, \dot r =  r^{(7-p)/2}\, .
\end{equation}
Just as in
\cite{Kabat:1999yq,Silverstein:2003hf,Alishahiha:2004eh},
causality in the supergravity dual implies a deceleration
mechanism. The VEV takes infinite time to roll to the singularity
at the origin of the Higgs branch, provided that the Coulomb
branch modulus $M$ vanishes.  We emphasize that we can not be
completely confident in the relation (\ref{decel}), because our
solution is rather far from self dual.  One way to verify this
sort of behavior would be to computate the higher derivative terms
in the effective action on the moduli space by integrating out the
non self-dual modes in the Dirac-Born-Infeld action.   This would
give a more precise result, since one could accurately identify
dual points on the Higgs branch.

By introducing a potential and coupling to (four dimensional)
gravity, it may be possible to obtain inflating cosmologies
similar to those discussed in
\cite{Silverstein:2003hf,Alishahiha:2004eh}, except that the
inflaton is a Higgs branch scalar rather than Coulomb branch
scalar.  The slow roll conditions would then be satisfied by
virtue of of the same sort of deceleration mechanism, which
follows from causality in the dual description.

\subsection{Rolling Higgs fields for non-zero
Coulomb branch VEVs}

Let us now consider the case in which the Coulomb-branch modulus
$M$ is non-vanishing.  Despite the insensitivity of the Higgs
branch metric to vector multiplet moduli,  motion towards a
singularity in the Higgs component of the moduli space depends on
$M$.

For instanton sizes large compared to $M$, the time evolution
should be approximately the same as for $M=0$.  However, for
sufficiently small instantons, with support in the region $y^my^m
<< M^2$,  the relevant part of the Dp+4 geometry is almost flat
and the action (\ref{unusualaction}) becomes;
\begin{align}\label{flac}
S \thickapprox \int dt\,d^4y\, {\rm tr}\left[F_{0m}F_{0m} -
\frac{1}{2d_p}M^{7-p}F^-_{mn}F^-_{mn}
 \right]
\end{align}
After a rescaling of coordinates to remove the factor
$M^{7-p}/d_p$ the equations of motion are the same as those of
Yang-Mills theory in flat five-dimensional space\footnote{The
$F^*F$ term now has constant coefficient and therefore does not
effect the equations of motion.}. There is no longer a causal
argument to prohibit the instanton from collapsing in a finite
time; a zero size instanton is far from the horizon, since $y^my^m
=0$ corresponds to $r^2 = M^2$.

After rescaling to remove the factor of $M^{7-p}/d_p$ in
(\ref{flac}), the equations of motion for the ansatz
(\ref{ansatz}), are
\begin{align}\label{EOMflat}
\left(-\partial_t^2  +
\partial_{\upsilon}^2 +
\frac{1}{\upsilon}\partial_{\upsilon} \right)f +
\frac{1}{\upsilon^2}(-4f + 6 f^2 -2f^3) = 0\, .
\end{align}
These equations have been studied in \cite{Bizon:2003kz} using a
perturbative expansion appropriate for solutions sufficiently
close to the static instanton solution;
\begin{align}
f(r,t) = \frac{2r^2}{r^2 + \rho(t)^2} + \epsilon(r,t) \, .
\end{align}
where $\epsilon$ is orthogonal (with respect to a certain inner
product) to the zero mode $\partial_\rho \left(\frac{2r^2}{r^2+
\rho(t)^2}\right)$ and has an expansion in time derivatives of
$\rho$. For collapsing initial conditions, the authors of
\cite{Bizon:2003kz} found that the instanton collapses in finite
time:
\begin{align}
\rho(t) = \frac{2}{3}\frac{t^*-t}{\sqrt{-{\rm ln}(t^*-t)}}
\end{align}
where $t^*$ is the time of collapse.

Although the instanton collapses in finite time,  the rate of
collapse $\dot\rho$ vanishes at $t=t^*$.  Energy conservation then
requires that the kinetic energy associated with the time
dependence of the instanton size must be converted into energy of
modes transverse to the moduli space, i.e. modes contained in
$\epsilon(r,t)$.  In the dual gauge theory,  the kinetic energy of
the rolling Higgs field is converted entirely into the energy of
particles which become light and are radiated during the approach
to the Higgs branch singularity. Assuming that inflating
cosmologies like those of
\cite{Silverstein:2003hf,Alishahiha:2004eh} can be realized during
the part of collapse for which the Higgs VEV (instanton size) is
large compared to $M$, then the part of the collapse for which the
Higgs VEV is small compared to $M$ seems a natural candidate for
reheating. We leave the problem of  constructing explicit
cosmologies based on strongly coupled Higgs branch dynamics for
elsewhere.

\section{Conclusions and Open Questions}

We have described properties of the supergravity background dual
to parts of the mixed Coulomb-Higgs branch of eight supercharge
Yang-Mills theories with fundamental representations. One outcome
of this work is a new set of constraints on unknown terms in the
non-Abelian Dirac-Born-Infeld action in a curved background, which
follow from the one to one correspondence between instantons and
the Higgs branch, together with the non-renormalization of the
metric on the Higgs branch. These constraints can be used to check
future calculations of higher dimension operators in the
Dirac-Born-Infeld action or, perhaps, to actually determine these
terms. The latter possibility will require additional information,
such as supersymmetry, since the number of constraints arising
from the Higgs branch-Instanton correspondence is fewer than the
number of unknown terms.

Without knowing the higher dimension operators in the
Dirac-Born-Infeld action, one can still study the low energy
dynamics of the Higgs branch at leading order in a large 't Hooft
coupling expansion. Using the supergravity description of the
Higgs branch, we have shown that the geodesic, or ``moduli
space,'' approximation is not suitable to describe the dynamics of
a Higgs expectation value rolling towards a singularity in the
moduli space.  In violation of predictions of the geodesic
approximation, this process is sensitive to the Coulomb-branch
moduli and exhibits deceleration.

We have demonstrated qualitative properties of motion on the Higgs
branch by studying time dependent solutions of the
Dirac-Born-Infeld equations of motion with instanton number one.
For vanishing Coulomb-branch moduli,  our analysis was imprecise,
since the time dependent solutions we discussed were far from self
dual.  Such solutions are not classified by the ADHM construction
and do not map in any obvious way to points on the Higgs branch. A
more precise analysis would require integrating out modes in the
Dirac-Born-Infeld action which are transverse to the instanton
moduli space, in order to obtain the higher derivative effective
action on the Higgs branch. At leading order in the large $N$
expansion, this amounts to summing tree level graphs.

Although the theories we have considered are in a Higgs phase,
computing the effective action on the Higgs branch might also be
an enlightening warmup for computing higher derivative terms of
the chiral Lagrangian in a yet to be constructed string dual of
QCD.  Like QCD in the chiral limit, the theories we consider have
a non-trivial moduli space.  At present, supergravity backgrounds
of ``QCD-like'' theories which have a massless meson akin to the
$\eta'$ at $N\rightarrow\infty$ have been found
\cite{Babington:2003vm}--\cite{Ghoroku:2004sp}, although a
holographically dual description of a theory with a spontaneously
broken non-Abelian chiral flavor symmetry remains elusive.

The deceleration mechanisms we have found are potentially
interesting from a cosmological point of view since, as pointed
out  in the context of Coulomb-branch dynamics
\cite{Silverstein:2003hf,Alishahiha:2004eh}, deceleration can give
slow roll inflation even when a steep potential is introduced. It
would be interesting to see what sort of cosmological models arise
from rolling Higgs moduli at strong coupling and large $N$. The
results described here are a preliminary step in investigating
this question.  For this purpose,  it would also be useful to know
the higher derivative effective action on the Higgs branch. It is
important to bear in mind that the effective action on the Higgs
branch is not reliable in situations in which there is production
of W-bosons or their superpartners which become light near the
origin of the Higgs branch. We have argued that this does in fact
occur for certain fixed Coulomb-branch moduli as one rolls to the
origin of the Higgs branch. Determining the details of the
particle production would require further analysis of the collapse
discussed in \cite{Bizon:2003kz}.  More generally,  a better
understanding of particle production in time dependent processes
for which field strengths on the Dp+4-brane are not self-dual is
desirable.

In the above discussion,  we have always taken the number of
Dp+4-branes and dissolved Dp-branes to be fixed in the
$N\rightarrow\infty$ limit, so that their backreaction can be
neglected.  The number of dissolved D3-branes has also been kept
fixed in this limit, for the same reason.  It would be interesting
to relax this constraint.  The ADHM constraints should also
classify the supergravity solutions associated with the Higgs
branch of the Dp-Dp+4 system in which all branes have been
replaced with geometry.  At present the fully localized
supergravity solutions associated with the D2-D6
\cite{Cherkis:2002ir,Erdmenger:2004dk} and D3-D7
\cite{Burrington:2004id} system have been studied, although not on
the Higgs branch.

\section{Acknowledgements}

I am especially grateful to S. Kovacs and B. Kulik, for
participation in early stages of this work.  I have also profited
from conversations with J. Erdmenger, K. Peeters, M. Schmaltz and
M. Zamaklar.

\section{Appendix: a few useful relations for 't Hooft tensors}

The 't Hooft tensor is defined by
\begin{align} \eta^a_{mn} =
-\eta^a_{nm},\, \eta^a_{ij} = \epsilon_{aij},\, \eta^a_{i4} =
\delta_{ia},
\end{align} where \begin{align}m,n=1\cdots 4,\qquad i,j=1\cdots 3,\qquad a
=1\cdots 3\, .\end{align} It satisfies
\begin{align}
&\eta^a_{mn} = \frac{1}{2}\epsilon_{mnrs}\eta^a_{rs}\\
&\eta^b_{ml}\eta^c_{nm}\epsilon_{bca}= -2\eta^a_{ln}\\
&\eta^b_{ml}\eta^b_{ms}=3\delta_{ls}\\
&\eta^e_{ml}\eta^e_{nt}=\delta_{mn}\delta_{lt} -
\delta_{mt}\delta_{ln}\\
&\epsilon_{abc}\epsilon_{dec}\eta^b_{ml}\eta^d_{ms}\eta^e_{nt} =
\eta^a_{ns}\delta_{lt} -\delta_{ln}\eta^a_{ts}
-3\delta_{ls}\eta^a_{nt}\\
&\epsilon_{pqmn}\eta^a_{nl} + \epsilon_{pqln}\eta^a_{nm}
=-2\epsilon^{abc}\eta^b_{pl}\eta^c_{qm}
\end{align}


\begin{thebibliography}{99}

\bibitem{Maldacena:1997re}
J.~M.~Maldacena, ``The large N limit of superconformal field
theories and supergravity,'' Adv.\ Theor.\ Math.\ Phys.\  {\bf 2},
231 (1998) [Int.\ J.\ Theor.\ Phys.\  {\bf 38}, 1113 (1999)]
[arXiv:hep-th/9711200].

\bibitem{Gubser:1998bc}
S.~S.~Gubser, I.~R.~Klebanov and A.~M.~Polyakov, ``Gauge theory
correlators from non-critical string theory,'' Phys.\ Lett.\ B
{\bf 428}, 105 (1998) [arXiv:hep-th/9802109].

\bibitem{Witten:1998qj}
E.~Witten, ``Anti-de Sitter space and holography,'' Adv.\ Theor.\
Math.\ Phys.\  {\bf 2}, 253 (1998) [arXiv:hep-th/9802150].

\bibitem{Guralnik:2004ve}
Z.~Guralnik, S.~Kovacs and B.~Kulik, ``Holography and the Higgs
branch of N = 2 SYM theories,'' arXiv:hep-th/0405127.

\bibitem{Argyres:1996eh}
P.~C.~Argyres, M.~R.~Plesser and N.~Seiberg, ``The Moduli Space of
N=2 SUSY {QCD} and Duality in N=1 SUSY {QCD},'' Nucl.\ Phys.\ B
{\bf 471}, 159 (1996) [arXiv:hep-th/9603042].

\bibitem{Kabat:1999yq}
D.~Kabat and G.~Lifschytz, ``Gauge theory origins of supergravity
causal structure,'' JHEP {\bf 9905}, 005 (1999)
[arXiv:hep-th/9902073].

\bibitem{Silverstein:2003hf}
E.~Silverstein and D.~Tong, ``Scalar speed limits and cosmology:
Acceleration from D-cceleration,'' arXiv:hep-th/0310221.

\bibitem{Alishahiha:2004eh}
M.~Alishahiha, E.~Silverstein and D.~Tong, ``DBI in the sky,''
arXiv:hep-th/0404084.

\bibitem{Chen}
X.~g.~Chen, ``Multi-throat brane inflation,'' Phys.\ Rev.\ D {\bf
71}, 063506 (2005) [arXiv:hep-th/0408084].

\bibitem{Fayyazuddin:1998fb}
A.~Fayyazuddin and M.~Spalinski, ``Large N superconformal gauge
theories and supergravity orientifolds,'' Nucl.\ Phys.\ B {\bf
535}, 219 (1998) [arXiv:hep-th/9805096].

\bibitem{Aharony:1998xz}
O.~Aharony, A.~Fayyazuddin and J.~M.~Maldacena,
 ``The large N limit of N = 2,1 field theories from three-branes in
F-theory,'' JHEP {\bf 9807}, 013 (1998) [arXiv:hep-th/9806159].

\bibitem{ADHM}
M.~F.~Atiyah, N.~J.~Hitchin, V.~G.~Drinfeld and Y.~I.~Manin,
``Construction Of Instantons,'' Phys.\ Lett.\ A {\bf 65}, 185
(1978).

\bibitem{Douglas:1996uz}
M.~R.~Douglas, ``Gauge Fields and D-branes,'' J.\ Geom.\ Phys.\
{\bf 28}, 255 (1998) [arXiv:hep-th/9604198].

\bibitem{Witten:1995gx}
E.~Witten, ``Small Instantons in String Theory,'' Nucl.\ Phys.\ B
{\bf 460}, 541 (1996) [arXiv:hep-th/9511030].

\bibitem{Dorey:2002ik}
N.~Dorey, T.~J.~Hollowood, V.~V.~Khoze and M.~P.~Mattis, ``The
calculus of many instantons,'' Phys.\ Rept.\  {\bf 371}, 231
(2002) [arXiv:hep-th/0206063].

\bibitem{Karch:2002sh}
A.~Karch and E.~Katz, ``Adding flavor to AdS/CFT,'' JHEP {\bf
0206}, 043 (2002) [arXiv:hep-th/0205236].

\bibitem{Douglas:1995bn}
M.~R.~Douglas, ``Branes within branes,'' arXiv:hep-th/9512077.

\bibitem{Itzhaki:1998dd}
N.~Itzhaki, J.~M.~Maldacena, J.~Sonnenschein and S.~Yankielowicz,
``Supergravity and the large N limit of theories with sixteen
supercharges,'' Phys.\ Rev.\ D {\bf 58}, 046004 (1998)
[arXiv:hep-th/9802042].

\bibitem{Skenderis:2002vf}
K.~Skenderis and M.~Taylor, ``Branes in AdS and pp-wave
spacetimes,'' JHEP {\bf 0206}, 025 (2002) [arXiv:hep-th/0204054].

\bibitem{Green:1996dd}
M.~B.~Green, J.~A.~Harvey and G.~W.~Moore, ``I-brane inflow and
anomalous couplings on D-branes,'' Class.\ Quant.\ Grav.\  {\bf
14}, 47 (1997) [arXiv:hep-th/9605033].

\bibitem{Wijnholt:2003pw}
M.~Wijnholt, ``On curvature-squared corrections for D-brane
actions,'' arXiv:hep-th/0301029.

\bibitem{Frey:2003jq}
A.~R.~Frey, ``String theoretic bounds on Lorentz-violating warped
compactification,'' JHEP {\bf 0304}, 012 (2003)
[arXiv:hep-th/0301189].

\bibitem{Koerber:2002zb}
P.~Koerber and A.~Sevrin, ``The non-abelian D-brane effective
action through order alpha'**4,'' JHEP {\bf 0210}, 046 (2002)
[arXiv:hep-th/0208044].

\bibitem{Gross:1986iv}
D.~J.~Gross and E.~Witten, ``Superstring Modifications Of
Einstein's Equations,'' Nucl.\ Phys.\ B {\bf 277}, 1 (1986).

\bibitem{Tseytlin:1986ti}
A.~A.~Tseytlin, ``Vector Field Effective Action In The Open
Superstring Theory,'' Nucl.\ Phys.\ B {\bf 276}, 391 (1986)
[Erratum-ibid.\ B {\bf 291}, 876 (1987)].

\bibitem{Tseytlin:1997cs}
A.~A.~Tseytlin, ``On non-abelian generalisation of the Born-Infeld
action in string  theory,'' Nucl.\ Phys.\ B {\bf 501}, 41 (1997)
[arXiv:hep-th/9701125].

\bibitem{Bergshoeff:2001dc}
E.~A.~Bergshoeff, A.~Bilal, M.~de Roo and A.~Sevrin,
``Supersymmetric non-abelian Born-Infeld revisited,'' JHEP {\bf
0107}, 029 (2001) [arXiv:hep-th/0105274].

\bibitem{Manton:1981mp}
N.~S.~Manton, ``A Remark On The Scattering Of Bps Monopoles,''
Phys.\ Lett.\ B {\bf 110}, 54 (1982).

\bibitem{Bizon:2003kz}
P.~Bizon, Y.~N.~Ovchinnikov and I.~M.~Sigal, ``Collapse of an
instanton,'' arXiv:math-ph/0307026.

\bibitem{Babington:2003vm}
J.~Babington, J.~Erdmenger, N.~J.~Evans, Z.~Guralnik and
I.~Kirsch, ``Chiral symmetry breaking and pions in
non-supersymmetric gauge /  gravity duals,'' Phys.\ Rev.\ D {\bf
69}, 066007 (2004) [arXiv:hep-th/0306018].

\bibitem{Babington:2003up}
J.~Babington, J.~Erdmenger, N.~J.~Evans, Z.~Guralnik and
I.~Kirsch, ``A gravity dual of chiral symmetry breaking,''
Fortsch.\ Phys.\  {\bf 52}, 578 (2004) [arXiv:hep-th/0312263].

\bibitem{Kruczenski:2003uq}
M.~Kruczenski, D.~Mateos, R.~C.~Myers and D.~J.~Winters, ``Towards
a holographic dual of large-N(c) QCD,'' JHEP {\bf 0405}, 041
(2004) [arXiv:hep-th/0311270].

\bibitem{Barbon:2004dq}
J.~L.~F.~Barbon, C.~Hoyos, D.~Mateos and R.~C.~Myers, ``The
holographic life of the eta','' JHEP {\bf 0410}, 029 (2004)
[arXiv:hep-th/0404260].

\bibitem{Evans:2004ia}
N.~J.~Evans and J.~P.~Shock, ``Chiral dynamics from AdS space,''
Phys.\ Rev.\ D {\bf 70}, 046002 (2004) [arXiv:hep-th/0403279].

\bibitem{Ghoroku:2004sp}
K.~Ghoroku and M.~Yahiro, ``Chiral symmetry breaking driven by
dilaton,'' Phys.\ Lett.\ B {\bf 604}, 235 (2004)
[arXiv:hep-th/0408040].

\bibitem{Cherkis:2002ir}
S.~A.~Cherkis and A.~Hashimoto, ``Supergravity solution of
intersecting branes and AdS/CFT with flavor,'' JHEP {\bf 0211},
036 (2002) [arXiv:hep-th/0210105].

\bibitem{Erdmenger:2004dk}
J.~Erdmenger and I.~Kirsch, ``Mesons in gauge / gravity dual with
large number of fundamental fields,'' arXiv:hep-th/0408113.

\bibitem{Burrington:2004id}
B.~A.~Burrington, J.~T.~Liu, L.~A.~Pando Zayas and D.~Vaman,
``Holographic duals of flavored N = 1 super Yang-Mills: Beyond the
probe approximation,'' arXiv:hep-th/0406207.



\end{thebibliography}
\end{document}